\documentclass[11pt]{article}

\usepackage{newtxtext,newtxmath}
\usepackage{graphicx}

\usepackage[letterpaper,margin=1in]{geometry}

\usepackage[colorlinks,linkcolor=blue,anchorcolor=blue,citecolor=blue,urlcolor=blue]{hyperref}

\renewenvironment{abstract}
    {\quotation}
    {\endquotation}

\date{}

\makeatletter
\renewcommand{\fnum@figure}{\textbf{Figure \thefigure}}
\renewcommand{\fnum@table}{\textbf{Table \thetable}}
\makeatother

\usepackage{scicite}

\usepackage{url}
\usepackage{lineno}
\usepackage{makecell}
\usepackage{indentfirst}

\usepackage{mathabx}
\usepackage{diagbox}

\def\scititle{Probing the Moon from Future Asteroid Impacts: A Review}
\title{\bfseries \boldmath \Large \scititle}

\author{
    Yifei Jiao$^{1}$,
    Bin Cheng$^{1\ast}$,
    Hexi Baoyin$^{1,2\ast}$\and
    \small$^{1}$Tsinghua University, Beijing, 100084, China.\\
    \small$^{2}$Inner Mongolia University of Technology, Inner Mongolia, 010051, China.\\
    \small$^\ast$Email: bincheng@tsinghua.edu.cn; baoyin@tsinghua.edu.cn\and
}

\begin{document} 

% Insert the title and author list
\maketitle
%\linenumbers
% Abstract, in bold
% There are strict length limits, and not all formats have abstracts.
% Consult the journal instructions to authors for details.
% Do not cite any references in the abstract.
\begin{abstract} \bfseries \boldmath
% Start with one or two sentences of background
% Then summarise the results of your observations, experiments, simulations etc.
% End with a statement of your main conclusions
The Earth--Moon system has been experiencing impacts from asteroids and comets for billions of years. The Moon, as an airless body, has preserved a distinct record of these events in form of impact craters and ejecta deposits, offering valuable insights into the impact history and surface evolution of the Moon.
However, the ancient impact relics can only provide limited information to the lunar interior structure, with an absence of the Moon's immediately dynamic response to the impact events.
With human lunar exploration entering an era of sustained presence, e.g., lunar research station and human return, we propose a new concept to probe the lunar subsurface and interior by investigating the future asteroid impacts in real time.
This can be achieved by integrating 1) ground- and space-based telescopes, 2) lunar-based seismometers and rovers, and 3) in-situ investigations around the impact sites.
A promising opportunity arises with the possible lunar impact of the 60-m-sized asteroid 2024~YR4 in 2032 (impact probability $\sim$4.3\%), an event of once-in-ten-thousand-years rarity.
Comprehensive observations of such events would greatly enhance our understanding of the Moon's structure and evolution, and have significant implications for the future lunar resource utilization and planetary defense missions.
\end{abstract}

\centerline{Keywords: near-Earth asteroids; lunar impact; impact scaling law; deep space exploration}

\subsection{Introduction}

The Moon is the only natural and long-standing satellite of our planet since the early solar system.
It have been widely thought that the Earth--Moon system are formed from a giant impact between the proto-Earth and a Mars-sized impactor Theia \cite{canup2001origin,yuan2023moon}, which explains the similarity between the Moon and the Earth's mantle, as well as the Moon's anomalously large mass and angular momentum \cite{asphaug2014impact}.
After the Earth--Moon formation, they could have experienced an intense bombardment period from about 4.1~Ga to 3.5~Ga, called the late heavy bombardment (LHB), forming about $95\%$ of all craters across the lunar surface \cite{gomes2005origin}.
Dynamical models suggest that the LHB impactors comprise both post-accretion and planetary-instability-driven populations \cite{bottke2017late}.
Then, over the past 3~Gyr, the lunar impact flux has remained relatively constant, with the near-Earth asteroids (NEAs) dominating the impactor population \cite{neukum2001cratering,xiao2024impact}.
As an airless body, the Moon has preserved a clear record of these historical impacts on its surface \cite{melosh1989impact}, from the ancient, global scale South Pole--Aitken impact basin \cite{su2025south,joy2025evidence}, to the fresh, tens-of-meter-diameter craters that formed annually \cite{speyerer2016quantifying}.
The lunar impact relics has provided invaluable insights for understanding the Moon's surface chronology \cite{stoffler2006cratering}, mare--highland dichotomy \cite{jones2022south}, and crustal and upper mantle structures \cite{lemelin2019compositions}.
Furthermore, seismic investigations by the Apollo missions \cite{nakamura1982apollo,garcia2019lunar}, combined with the gravitational and topographic data from the Gravity Recovery and Interior Laboratory (GRAIL) mission \cite{williams2014lunar}, have revealed a clearly layered, crust--mantle--core interior of the Moon.

\begin{figure}[b!]
    \centering
    \includegraphics[width=0.95\textwidth]{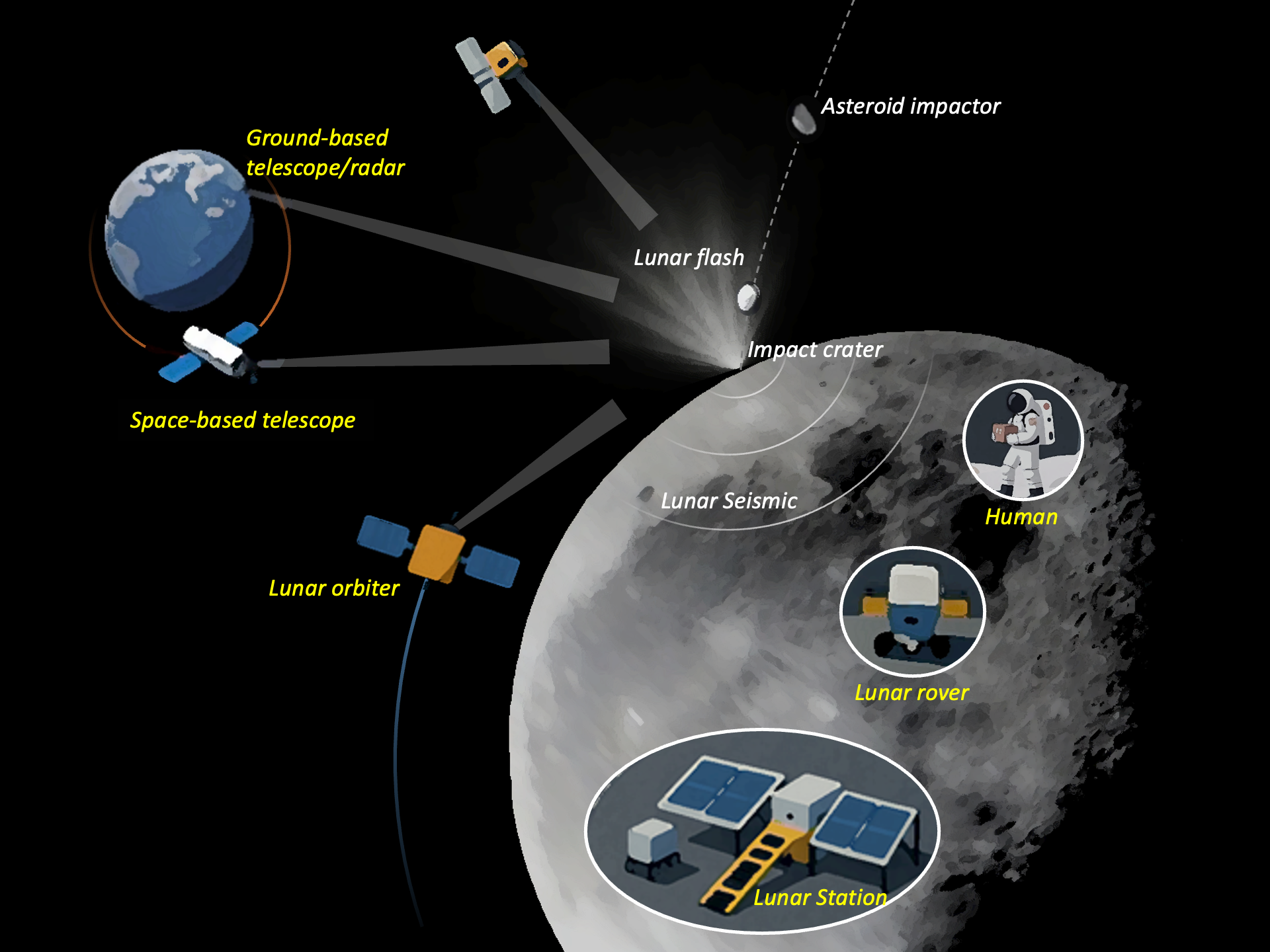}
    \\[-10pt]
    \caption{\textbf{Scientific opportunities during an asteroid impact with the Moon.}
    }
    \label{fig:1}
\end{figure}

However, the detailed structure beneath the lunar surface remains poorly understood, due to 1) the limited resolution and coverage of existing seismic and gravity data, and 2) the lack of dynamic response information of the Moon to external forces, e.g., asteroid impacts.
The same situation applies to Mars, until the recent observations of Martian impacts with the co-investigation of orbital imaging and surface seismology \cite{posiolova2022largest}.
Such large impacts on Mars (magnitudes greater than 4, craters larger than 100~m) have greatly improved our understanding of its subsurface composition and internal structure \cite{lognonne2023mars}.
This arise a promising opportunity for the future lunar exploration to conduct systematical investigations of asteroid impacts on the Moon (to distinguish from smaller scale meteoroid impacts, we call it an asteroid impact if the impactor size is larger than 1~m).
In this paper, we present an overview of the current knowledge of asteroid impacts on the Moon, and propose the future scientific opportunities to investigate the dynamic response of the Moon during such impacts (as illustrated in Figure~\ref{fig:1}).
This would greatly enhance our understanding of the formation and evolution of the Moon itself and the Earth--Moon system.

\subsection{Asteroid impacts on the Moon}

\subsubsection{Asteroid impactor flux}

Most asteroids have settled in the main asteroid belt since billions of years ago \cite{morbidelli2015dynamical}.
During the long-term dynamical evolution under gravitational perturbations of giant planets \cite{morbidelli2002origin} and the Yarkovsky effect \cite{bottke2006yarkovsky}, some of the main belt objects would migrate inwards to become NEAs and possibly collide with the Earth--Moon system \cite{granvik2018debiased}.
Besides these NEAs, the Moon also experiences impacts from comets \cite{bell1987recent}, e.g., Jupiter family comets and long period comets, which are characterized by their high-porosity, low-density materials and higher impact velocities.
Since the cometary impacts contribute only less than $1\%$ compared to NEAs \cite{yeomans2013comparing}, they are not considered in this paper.

As for the impactor flux, according to Neukum's lunar crater production model \cite{neukum2001cratering}, there would be on average one 50-m-diameter crater formed on the Moon per year, as also seen by the Lunar Reconnaissance Orbiter (LRO) from 2009 to 2015 \cite{speyerer2016quantifying}.
We can then estimate the impactor population using the impact scaling laws (Section~2.2), e.g., a 50-m-diameter lunar crater would correspond to a 1.5-m rocky impactor, assuming the typical impact velocity of 15~km/s \cite{yue2013projectile} and impact angle of 45$^\circ$ \cite{shoemaker1962interpretation}.
On the other hand, the Earth impact flux has been recorded as fireballs in Earth's atmosphere \cite{brown2002flux}, so we can simply normalize it with the cross-section area ratio of the Moon and Earth
\begin{equation}
    N(d)=10^{1.568} d^{-2.7} \left(\frac{R_{\leftmoon}}{R_{\Earth}}\right)^{2}
\end{equation}
where $N(d)$ is the cumulative number of $>d$ impactors that hit the Moon per year; $d$ is in meter; $R_{\leftmoon}$ and $R_{\Earth}$ represent the radius of the Moon and Earth, respectively.
Overall, both the lunar crater production and Earth-based fireball records suggest a considerable amount of asteroid impacts upon the Moon, within a relatively short timespan.
As a reference, the average frequency of a 10-m lunar impactor is about 180 years; for a 60-m impactor, it would take over 23,000 years.

Then, have such lunar impact events been seen/recorded during human history, and what can we expect in the near future?
One of the most famous historic record is from the Medieval chronicles of Gervase of Canterbury: ``\textit{upper horn of a new moon split and from the division pointfire, hot coals, and sparks spewed out}'' in A.D. 1178 \cite{newton1972medieval}, which was thought to form the 22-km-diameter Giordano Bruno crater on lunar farside but has been ruled out now \cite{calame1978lunar,morota2009formation}.
Modern astronomical observations have recorded many more lunar flashes in the last few decades \cite{yanagisawa2006first,madiedo2018first}, most of which are caused by $<$1~m meteorite impacts.
However, it is not easy to say how many lunar impacts we will see in the near future, since such impactors are hard to detect due to their small size and dim brightness, as well as the limited capabilities of our telescopes.
Of all known asteroids in NASA JPL's Small-Body Database, we do find some candidates that might hit the Moon (with the minimum lunar distance of $<R_\leftmoon$ during a close approach) in the future.
As shown in Table~\ref{tab:1}, the asteroid 2024~YR4 would be the largest one of those, with an expected size of 60~m assuming an albedo of 0.15 \cite{rivkin2025jwst}, comparable size to the Tunguska impactor \cite{sekanina1998evidence}.
Using observations through May 2025, 2024~YR4 has a 4.3\% possibility to impact the Moon in 2032 and won't impact the Earth \cite{wiegert2025potential}.
Further observations will help to determine the precise orbit of 2024~YR4 when it goes back into the vicinity of Earth in 2028.
Whether the asteroid hits the Moon or makes a close flyby in 2032, it would be a once-in-ten-thousand-years event and definitely deserve further investigations.

\begin{table}[htb!] % Do NOT use \begin{table*}
\centering
\caption{\textbf{Possible asteroid impacts with the Moon in the next 50 years.} The close-approach data are from NASA JPL's Small-Body Database (https://ssd-api.jpl.nasa.gov) as of June 2025. The distances are measured from the center of the Moon, with the minimum and maximum values representing 3-$\sigma$ errors. $H$ is the absolute magnitude.}
\label{tab:1}
\newcolumntype{C}[1]{>{\centering\arraybackslash}p{#1}}
\begin{tabular}{ccC{2cm}C{2cm}C{2cm}C{1.3cm}c}
    \\[-10pt]
    \hline
    Designation & 
    \makecell{\vspace{-5pt}Close approach time\\(TDB)} & 
    \makecell{\vspace{-5pt}Nominal\\\vspace{-5pt}distance\\(au)} & 
    \makecell{\vspace{-5pt}Minimum\\\vspace{-5pt}distance\\(au)} & 
    \makecell{\vspace{-5pt}Maximum\\\vspace{-5pt}distance\\(au)} & 
    \makecell{\vspace{-5pt}Relative\\\vspace{-5pt}velocity\\(km/s)} & 
    $H$ \\
    \hline
    2019 EH1 & 2032-Feb-29 17:47 & $5.63\times10^{-4}$ & $3.34\times10^{-6}$ & $1.56\times10^{-1}$ & 13.9 & 30.1\\
    2024 YR4 & 2032-Dec-22 15:10 & $7.14\times10^{-5}$ & $6.78\times10^{-6}$ & $5.66\times10^{-4}$ & 13.9 & 23.9\\
    2024 JP17 & 2038-May-17 06:41 & $5.38\times10^{-3}$ & $1.17\times10^{-5}$ & $4.54\times10^{-2}$ & 11.6 & 26.8\\
    2006 JY26 & 2074-May-04 02:14 & $7.14\times10^{-3}$ & $2.14\times10^{-6}$ & $2.60\times10^{-2}$ & 3.8 & 28.4\\
    \hline
\end{tabular}
\end{table}

\subsubsection{Impact cratering}

The direct result of an asteroid impact on the Moon would be a crater at the impact site.
Previous works have employed theory models \cite{holsapple1993scaling}, numerical simulations \cite{davison2011numerical,jiao2024sph}, laboratory experiments \cite{pierazzo2000understanding}, and even artificial impacts on the Moon \cite{schultz2010lcross} to understand what happened during the impacts.
Here we mainly use theoretical and empirical relations to provide an order-of-magnitude estimate of the impact results.
Using the point-source theory \cite{holsapple1993scaling}, the impactor can be characterized with its radius $a$, velocity $U$ (for oblique impacts, we can simply use the vertical component of the velocity \cite{housen2011ejecta}), and density $\delta$. The target properties include the density $\rho$, lunar gravity of $g=1.625$~m/s$^2$, and the strength measure $Y$.
Then, the crater size (volume of $V$) can be expressed in a general form of
\begin{equation}
    \pi_V = K_1\left\{\pi_2\left(\frac{\rho}{\delta}\right)^{\frac{6v - 2 - \mu}{3\mu}} + K_2\left[\pi_3\left(\frac{\rho}{\delta}\right)^{\frac{6v - 2}{3\mu}}\right]^{\frac{2 + \mu}{2}}\right\} ^{-\frac{3\mu}{2+\mu}}
    \label{eq:scaling}
\end{equation}
where the dimensionless terms are
\begin{equation}
    \pi_V = \frac{\rho V}{m}, \quad \pi_2 = \frac{ga}{U^2}, \quad \pi_3 = \frac{Y}{\rho U^2}
\end{equation}
where $m=\frac{4}{3}\uppi \delta a^3$ is the impactor mass.
The crater radius is then determined as $R=K_RV^{\frac{1}{3}}$, with $K_R=1.3$ for the typical bowl-shaped, simple craters \cite{holsapple2022main}.
When the gravity term $\pi_2$ dominates the crater scale and the strength term $\pi_3$ can be ignored, it is called the gravity regime; conversely, the strength regime occurs.
The strength--gravity regime transition happens when $\pi_2=(\pi_3)^{\frac{2+\mu}{2}}$.
For lunar surficial regolith, the constants in Eq.~(\ref{eq:scaling}) are typically taken as $K_1=0.132$, $K_2=0.26$, $\mu=0.41$, and $\nu=0.33$ \cite{holsapple1993scaling}, the regolith strength is adopted as $Y=10~{\rm kPa}$, and the bulk density is $\rho=1,500~{\rm kg\,m^{-3}}$ \cite{mitchell1972mechanical}.
Assuming a typical impact velocity of 15~km/s \cite{yue2013projectile} and a rocky impactor density of $\delta=3,000~{\rm kg\,m^{-3}}$, we estimate the transition size of about 0.1~m for the impactor.
This suggests that all asteroid impacts discussed in this paper are dominated by gravity regime.

Considering an 10-m impactor at 15~km/s, the produced crater would be about 200~m in diameter; for 2024~YR4, the expected crater would be about 1,000~m wide.
The depth-to-diameter ratios of craters in comparable sizes can vary from $\sim$0.2 to $\sim$0.1 \cite{sun2018investigation}, generally shallower than that of larger craters.
Since the average depth of the lunar regolith is only about 5~m for mare and 12~m for highland \cite{shkuratov2001regolith}, the impact would definitely excavate and expose the subsurface materials beneath the lunar regolith.
The crater shape is generally circular rather than elliptical, as long as the impact angle is not very oblique, with the threshold of $\sim$20$^\circ$ from surface \cite{collins2011size}.
While the ejecta pattern can be quite asymmetric in oblique cases \cite{luo2022ejecta}.
During the cratering process, most of the impact energy is convert into two parts: 1) the internal energy, which can heat, melt, and even vaporize the materials \cite{cintala1992impact}, and 2) the kinetic energy of the ejecta, which is further controlled by the lunar gravity (Section~2.4).
Therefore, the final crater may exhibit a shallow, circular morphology, surrounded by a thin ejecta blanket and containing thermally altered and/or fresh subsurface materials on its floor.
This could be further investigated using remote and in-situ techniques.

\subsubsection{Lunar impact flash}

Immediately after the asteroid impact, the shock wave would produce an expanding gas-plasma cloud, followed with a cooling phase that lead to condensation of liquid droplets \cite{ait2015first}.
The expansion and cooling of the clouds would emit significant radiation that can be observed as lunar flashes.
The radiated energy can be simply estimated to be proportional to the kinetic energy of the impactor, $\eta E_{\rm imp}$, where the luminous efficiency $\eta$ is depending on the impact velocity and typically taken as $2\times10^{-3}$.
Then, to derive the peak apparent magnitude $m_{\rm V}$ assuming an observer at $d_{\rm obs}$, we can numerically solve the following relation \cite{merisio2023present}
\begin{equation}
    \eta E_{\rm imp} = 0.1987 \tau P_0 = 0.1987 \tau \left(F_0 10^{-\frac{m_{\rm V}-m_0}{2.5}}f_{\rm a}\uppi d^2_{\rm obs} \Delta\lambda\right)
\end{equation}
where $\tau=(77.6\pm34.4){\rm e}^{-(0.94\pm0.06)m_{\rm V}}$ is the duration of the flash in seconds \cite{madiedo2015analysis}, $P_0$ is the peak radiated power as a function of $m_{\rm V}$ in watts, $F_0=1.36949\times10^{-10}$~W/m$^2$/m is the flux density of a reference source of magnitude $m_0=21.1$ \cite{madiedo2015analysis}, $f_a$ is the degree of anisotropy of the light emission ranging from 2 to 4, and $\Delta\lambda$ is the bandwith of the spectral domain (0.4--0.9~$\upmu$m).
For a 10-m impactor at 15~km/s, the impact flash would have a peak apparent magnitude of 0.1 with a duration of about 70~s; for 2024~YR4, the expected impact flash would have a peak magnitude of $-$2.7 and may last for 1,000~s, when observing from the Earth.

The spectral emissive power $L$ of an impact flash could be represented by the emission of a black body at a given effective temperature $T_{\rm e}$ \cite{bouley2012power}
\begin{equation}
    L(\lambda,T_{\rm e}) = \frac{2\uppi h c^2}{\lambda^5} \frac{1}{{\rm e}^{\frac{hc}{\lambda k_{\rm B} T_{\rm e}}}-1}
    \label{eq:black-body}
\end{equation}
where $L$ is in W/m$^2$/m, $\lambda$ is the wavelength, $c$, $h$ and $k_{\rm B}$ are respectively the light speed in vacuum, the Planck's constant and the Boltzman's constant.
The equivalent temperature $T_{\rm e}$ at the peak power $P_0$ can be solved from the integration of Eq.~(\ref{eq:black-body}) over the spectral bandwidth [$\lambda_1,\lambda_2$]
\begin{equation}
    \frac{P_0}{\beta \uppi R^2} = \int_{\lambda_1}^{\lambda_2} L(\lambda, T_{\rm e}) {\rm d}\lambda
\end{equation}
where $\beta \uppi R^2$ is the effective flash area, with the factor $\beta$ ranging from 4 to 25 \cite{merisio2023present}, and $R$ is the crater radius.
We can thus estimate the equivalent temperature of the radiating impact plume to range from 1850~K to 2350~K, which depends more on the $\beta$ factor rather than the impact energy.
With Eq.~(\ref{eq:black-body}), the peak emission of the impact flash is located at about 1.2--1.6~$\upmu$m (near infrared), with a peak power of $3\times10^{11}$--$9\times10^{11}$~W/m$^2$/m.

\subsubsection{Impact ejecta evolution}

Generally, the velocity and mass distributions of the impact-induced ejecta follow a simple power law: as goes from the point of impact to the crater edge, the eject velocity $v(r)$ decreases while the ejecta mass $M(r)$ accumulates
\begin{equation}
    \frac{v(r)}{U} = C_1\left[\frac{r}{a}\left(\frac{\rho}{\delta}\right)^\nu\right]^{-\frac{1}{\mu}}
    \label{eq:vr}
\end{equation}
\begin{equation}
    \frac{M(r)}{m} = \frac{3}{4\uppi} \frac{1}{K_R^3} \frac{\rho}{\delta} \left( \frac{r}{a} \right)^3
    \label{eq:mr}
\end{equation}
where $r$ denotes the distance from the impact site, following $n_1a\leq r \leq n_2R$, and $C_1$ is a constant with the typical value of 0.55 for sand \cite{housen2011ejecta}.
The constants $n_1$ and $n_2$ depend on the impact speed, projectile shape, materials, etc. \cite{housen2011ejecta}.
With Eqs.~(\ref{eq:vr}--\ref{eq:mr}), it is easy to obtain the cumulative linear momentum and the kinetic energy of the ejecta as $\int {\rm d}M v\cos\theta$ and $\int 0.5{\rm d}M v^2$, where $\theta$ is the assumed launch angle from the ground plane.
As a typical reference, the total momentum of the ejecta is approximately equal in magnitude but opposite in direction to that of the impactor, assuming symmetry.
While the kinetic energy of the ejecta is only about 10$\%$ of the impactor's.

Following the constants for lunar regolith in Section~2.2 and assuming an impact velocity of $U=15$~km/s, we can estimate the threshold of lunar escape ejecta with $v>2.38$~km/s.
According to Eqs.~(\ref{eq:vr}--\ref{eq:mr}), only the ejecta within $r/a<2.1$ can escape from the lunar gravity, and the total escape mass is about 0.5 times of the impactor mass.
Also note that the actual escape mass could be several times greater in oblique cases, such as 45$^\circ$ \cite{artemieva2008numerical}.
These high-speed ejecta would escape into geocentric orbits and even heliocentric space, most of which eventually hit the Earth and become lunar meteorites \cite{gladman1995dynamical}.
Some of the escape lunar debris could still survive in space as NEAs, with possible candidates such as the Earth's quasi-satellite 2016~HO3 \cite{jiao2024asteroid} and the temporally mini-moon 2024~PT5 \cite{kareta2025lunar}, which may come from some ancient, larger scale lunar impacts.
While the low-speed ejecta would fall back on the lunar surface in ballistic trajectories, forming bright rays and secondary craters around the primary crater \cite{singer2020lunar}.
The ground distance $l$ traveled by the ejecta is given by \cite{melosh1989impact}
\begin{equation}
    l = 2 R_{\leftmoon} \tan^{-1} \left(\frac{v^2\sin\theta\cos\theta}{R_{\leftmoon}g-v^2\cos^2\theta}\right)
    \label{eq:ballistic}
\end{equation}
where $\theta$ is the launch angle from the ground plane, with the typical value of 45$^\circ$.
Combining Eqs.~(\ref{eq:vr}--\ref{eq:ballistic}), we can derive a mass density of the fallen ejecta shower as a function of the ground distance from the impact point.
This would be helpful to study the regolith growth and secondary crater distribution on the lunar surface, as well as the impact risk analysis of the future lunar explorations.
Besides the marco-scale ejecta, there would also be a large amount of micron-sized lunar dust particles, which can be significantly affected by non-gravitational forces and have distinct dynamical fates \cite{yang2022review}.

\subsubsection{Impact-induced moonquake}

A moonquake is the lunar equivalent of an earthquake, and has been widely detected during the Apollo missions from 1969 to 1977.
There are different types of moonquakes: artificial/meteoroid impacts, shallow quakes, deep quakes, and thermal quakes, where the meteoroid impacts and deep quakes are more frequent while the shallow quakes are rare but have larger magnitudes \cite{nunn2020lunar}.
In asteroid impact scenarios, the impact-induced shock waves would gradually degrade into stress waves during propagation, and finally become elastic (seismic) waves through the target body.
The seismic energy $M_{\rm seismic}$ of an asteroid impact can be estimated from the Gutenberg--Richter magnitude--energy relation \cite{melosh1989impact}
\begin{equation}
    \log_{10}\left(kE_{\rm imp}\right) = 4.8 + 1.5M_{\rm seismic}
\end{equation}
where $k=10^{-4}$ is commonly used as the seismic efficiency, $E_{\rm imp}$ is the kinetic energy of the impactor in joule.
For example, a 10-m impactor, with impact velocity of 15 km/s, would cause a moonquake with a magnitude of 3.6; in the case of 2024~YR4, the seismic magnitude is estimated to be 5.1.

Impact-induced seismic waves typically include body waves---P-waves (primary, compressive) and S-waves (secondary, shear)---as well as the surface Rayleigh waves \cite{nunn2024artificial}.
Body waves can travel through the Moon's interior, except for S-waves, which cannot propagate through liquid regions such as the lunar core. In contrast, surface waves travel along the lunar surface.
Due to the Moon's low intrinsic attenuation in the interior and the strong scattering properties of the lunar regolith, the seismic waves can propagate with long durations and retain high frequencies \cite{garcia2019lunar}.
For a given impactor impulse $mU$, the acceleration spectral density $A$ of the body waves can be estimated using an empirical formula \cite{gudkova2011large}
\begin{equation}
    A_{\rm seismic} = 1.1\times10^{-12} S mU f^3 / D
\end{equation}
where $A_{\rm seismic}$ is in $\rm m/s^2/Hz^\frac{1}{2}$, $S$ is the seismic amplification (typical values of 1.7), $mU$ is the impactor impulse in $\rm kg\,m/s$, $f$ is the frequency in Hz, and $D$ is the epicentral distance in km.
Considering a 10-m asteroid impactor at 15~km/s, the spectral amplitude over a distance of 2,000~km is about $\rm 2\times10^{-5}~m/s^2/Hz^\frac{1}{2}$ at 1~Hz (typical frequency for body waves), or $\rm 2\times10^{-8}~m/s^2/Hz^\frac{1}{2}$ at 0.1~Hz (surface waves).
Compared with the Apollo noise detection level of $\rm 4\times10^{-10}~m/s^2/Hz^\frac{1}{2}$ \cite{gudkova2011large}, such impact seismic waves would be detectable with high signal-to-noise ratios (SNR) across the whole lunar surface, regardless of the shadow zones due to refraction or non-propagation at the liquid core.

According to all the theoretical and empirical models described in this section, Table~\ref{tab:2} summarizes the expected lunar impact results for asteroids of various sizes that could strike the Moon in the near future.

\begin{table}[htb!]
\centering
\caption{\textbf{Expected lunar impact results with different asteroid sizes.} For simplicity, we assume an impact velocity of 15~km/s and an impact angle of 45$^\circ$ in all cases. Note that the 60-m results are provided for general reference only, and do not correspond exactly to the 2024~YR4 case.}
\label{tab:2}
\newcolumntype{C}[1]{>{\centering\arraybackslash}p{#1}}
\begin{tabular}{lC{2cm}C{2cm}C{2cm}}
    \\[-10pt]
    \hline
    Impactor diameter (m) & 1.5 & 10 & 60 \\
    \hline
    Average time interval (yr) & 1 & 180 & 23,000 \\
    Impact kinetic energy (J) & $6.0\times10^{11}$ & $1.8\times10^{14}$ & $3.8\times10^{16}$ \\
    Escape ejecta mass (kg) & $2.6\times10^3$ & $7.8\times10^5$ & $1.7\times10^8$ \\
    Crater diameter (m) & 44 & 210 & 940 \\
    Flash visual magnitude & 3.1 & 0.1 & $-$2.8 \\
    Flash duration (s) & 4.1 & 72 & 1090 \\
    Seismic magnitude & 2.0 & 3.6 & 5.2 \\
    \hline
\end{tabular}
\end{table}

\subsection{Scientific opportunities}

Investigating future asteroid impacts on the Moon offers a unique opportunity to address several fundamental questions, including 1) the entire dynamical process of an asteroid impact on the Moon; 2) the impact-induced seismic response of the lunar interior; and 3) the freshly exposed lunar subsurface materials and the exogenous impactor remnants.
These questions can be investigated by combining ground- and space-based telescopes, lunar seismometers and rovers, and in-situ analyses near the impact sites (summarized in Table~\ref{tab:3}).
The resulting discoveries would significantly advance our understanding of the Moon's internal structure and surface evolution, as well as the impact history of the Earth--Moon system.

\newcommand{\circled}[1]{\textcircled{\raisebox{-1.2pt}{#1}}}
\begin{table}[htb!]
\centering
\caption{\textbf{The unsolved scientific questions and proposed investigations during a future lunar impact event.} The available information before, during, and after the impact process may include: \circled{1} the impactor's orbital and physical characteristics, \circled{2} impact flash, \circled{3} crater formation, \circled{4} ejecta evolution, \circled{5} impact-induced seismic waves, \circled{6} sampling of impact products.}
\label{tab:3}
\newcolumntype{C}[1]{>{\centering\arraybackslash}p{#1}}
\begin{tabular}{lccc}
    \\[-10pt]
    \hline
    \diagbox[width=15em,height=3em]{Science Questions}{Investigations} & Remote obs. & Lunar-based & In-situ \\
    \hline
    Q1: \makecell[lt]{\vspace{-5pt} Entire dynamical process of \\ lunar impacts} & \circled{1}\circled{2}\circled{3}\circled{4} & \circled{4}\circled{5} & \circled{3}\circled{6} \\
    Q2: \makecell[lt]{\vspace{-5pt} Lunar deep interior and its \\ dynamical response} & \circled{1} & \circled{5} & \circled{6} \\
    Q3: \makecell[lt]{\vspace{-5pt} Lunar subsurface materials \\ and exogenous remnants} & \circled{1}\circled{3}\circled{4} & \circled{4}\circled{5} & \circled{3}\circled{6} \\
    \hline
\end{tabular}
\end{table}

\subsubsection{Remote observations}

Prior studies of lunar impacts have primarily focused on the more common meteoroid-scale impacts.
These efforts typically fall into two categories: 1) continuous monitoring of the Moon's night side to detect impact flashes, such as those conducted by the NASA Lunar Impact Monitoring Program \cite{suggs2008nasa} and ESA's Near-Earth object Lunar Impacts and Optical TrAnsients (NELIOTA) project \cite{xilouris2018neliota}; and 2) post-impact identification of new craters using images from lunar orbiters, such as the LRO \cite{speyerer2016quantifying}.
However, the rare asteroid impacts on the Moon can sometimes be predicted in advance, such as the possible lunar impacts listed in Table~\ref{tab:1}.
Also note that the current detection completeness of small NEAs remains low: approximately $2\%$ for 50-meter-sized NEAs and only $0.1\%$ for 10-meter-sized ones \cite{harris2021population}.
Such small objects are typically detected only shortly, e.g., a few hours or days, before they approach the Earth--Moon system.
This may allow a quick, preliminary assessment of the impactor's orbital and physical characteristics.
Nevertheless, when predictions are available, they enable the use of large telescopes and radar systems to observe not only the impact flash and final craters, but also the entire impact process in real time, which includes 1) the optical and thermal emissions, 2) crater formation, and 3) ejecta/dust evolution.
The best resolution for a 10-m-class ground telescope, such as the Very Large Telescope (VLT), is expected to be a few tens of meters per pixel at lunar distance.
Earth-based radar systems, including the Goldstone Solar System Radar (GSSR) \cite{slade2010goldstone} and the China Compound Eye \cite{ding2024china}, can also contribute to characterizing the dynamic evolution of crater morphology during the impact process.
However, ground-based observations are unable to detect lunar impacts occurring on the farside or under unfavorable observation conditions.
Space-based telescopes---such as the Hubble Space Telescope \cite{savage1996interstellar}, James Webb Space Telescope (JWST) \cite{sabelhaus2004overview}, and the future Chinese Space Station Telescope (CSST) \cite{gong2019cosmology}---along with lunar orbiters (sub-meter resolution) such as the LRO \cite{speyerer2016quantifying} and the Chandrayaan-2 orbiter \cite{chowdhury2020imaging}, can provide detailed observations from different perspectives and across various wavelengths.

\subsubsection{Lunar-based investigations}

\begin{figure}[tb!]
    \centering
    \includegraphics[width=0.75\textwidth]{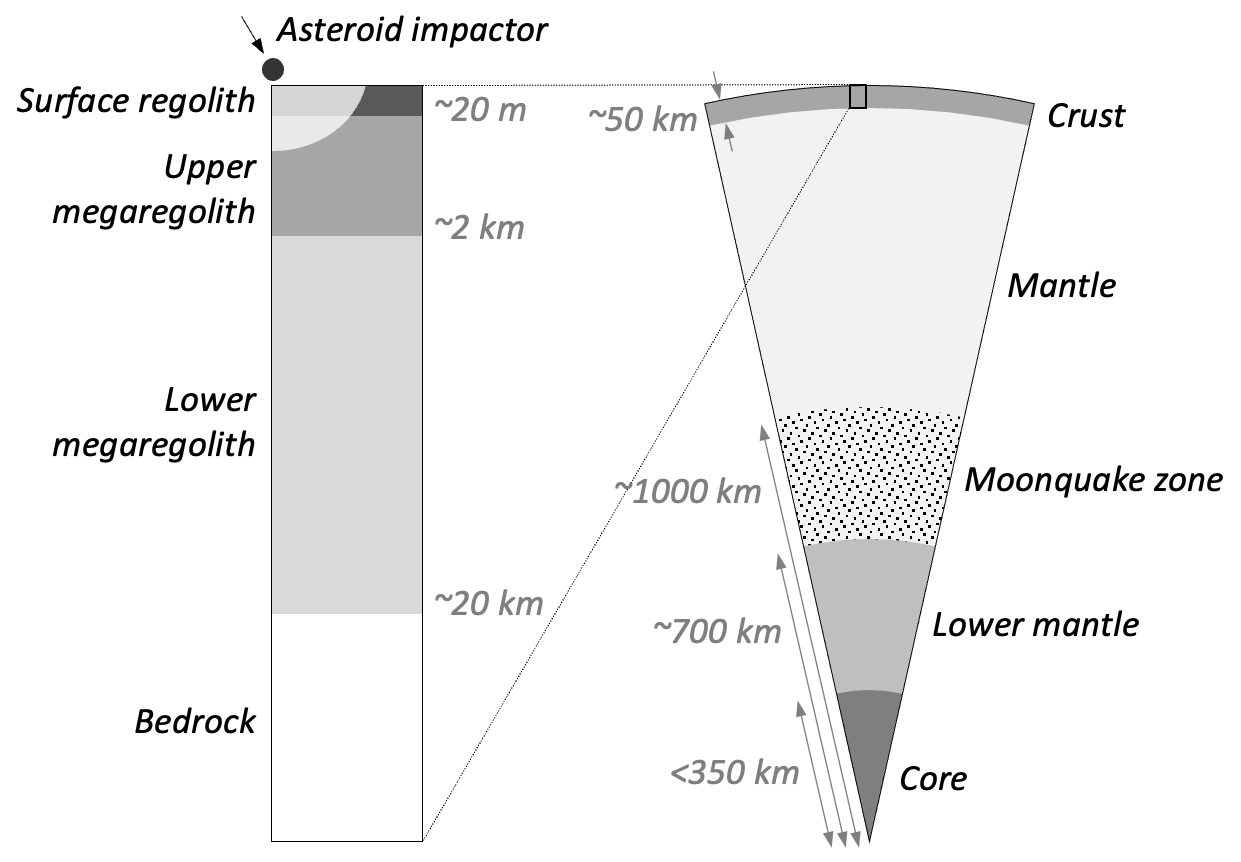}
    \\[-10pt]
    \caption{\textbf{Interior and crustal structure of the Moon \cite{nunn2020lunar,richardson2020modeling}.} For comparison, with a 10-m-diameter rocky impactor at 15~km/s and 45$^\circ$, the resulting lunar crater is expected to have a diameter of $\sim$200~m and a depth of $\sim$40~m.
    }
    \label{fig:2}
\end{figure}

Seismic data from the Moon was last collected during the Apollo missions in the 1970s \cite{nunn2020lunar}, and no active seismometers are currently operating on the lunar surface.
Fortunately, upcoming missions will re-initiate the lunar seismic monitoring.
The Chang'E~7 and Chang'E~8 missions of the Chinese Lunar Exploration Program (CLEP), scheduled for 2026 and 2028 respectively, will each deploy a lunar seismometer on the lander, which will explore the Moon's south pole \cite{wang2024scientific,he2025dual}.
The NASA's Artemis~III mission in 2027 will also carry a compact, autonomous seismometer suite, called the Lunar Environment Monitoring Station (LEMS), and land on the south polar region \cite{benna2024lunar}.
Furthermore, we would expect a multi-station seismic network that include at least 3--4 stations and cover the whole lunar surface \cite{yamada2011optimisation}, in order to investigate the three-dimensional lunar seismic activity during the future asteroid impacts.
The future lunar seismic investigations would help to further constrain 1) the lunar deep interior, i.e., the core and deep mantle structures (Figure~\ref{fig:2}); 2) the lunar crustal dichotomy between nearside/farside; and 3) the lunar regolith properties.

However, there is a new concern that the lunar exploring activities could be at risk due to impact-induced ejecta and moonquakes.
For the impact ejecta that would fall back onto the surface under lunar gravity, we can use the total kinetic energy as a damage measurement.
According to Sec.~2.4, the kinetic energy of the ejecta upon falling back should equal their kinetic energy at launch, which is estimated to be $\sim$2$\%$ of the impact energy $E_{\rm imp}$.
Using Eq.~(\ref{eq:ballistic}), we can further estimate the kinetic energy distribution of the fallen ejecta across the lunar surface, showing that approximately 75$\%$ of the total energy is deposited on the hemisphere facing the impactor.
Recall that the total post-impact kinetic energy is about 10$\%$ of $E_{\rm imp}$, suggesting that the high-speed escaped ejecta carries the majority of the kinetic energy.
In the case of 2024~YR4's possible impact with the Moon, the kinetic energy of the fallen ejecta would be about $7\times10^{14}$~J (150 kton TNTs equivalent).
This means that the whole lunar surface would experience a low-speed, short-duration ejecta shower that is comparable to the total meteoroid impact energy in 100 years.
As for the seismic damage, since the asteroid impacts primarily produce P-waves, which is not as destructive as S-waves from tectonic activities.
It has been suggested that an impact-generated seismic disturbance is equal in destructiveness to an earthquake one magnitude smaller \cite{melosh1989impact}.
The seismic magnitude of the possible 2024~YR4's impact is estimated to be 5.1, which could be comparable in effect to a magnitude 4.1 earthquake and less likely to cause serious damage to distant lunar-based facilities.

\subsubsection{In-situ explorations}

While it is risky to perform in-situ investigations during the impact process, a post-impact exploration would be both feasible and valuable.
The post-impact, in-situ investigations can be performed by the future lunar robotic rovers \cite{ye2017overview} and even human astronauts.
Since the Apollo~17 mission in 1972, the upcoming Artemis missions \cite{smith2020artemis} and the Chinese crewed lunar missions \cite{lin2024return} will return humans to the Moon by around 2030.
And the International Lunar Research Station is expected be utilized from 2036 \cite{xu2022brief}, providing an unique opportunity to support such in-situ explorations.
In the history of lunar exploration, no fresh lunar crater has been investigated so closely and immediately after its formation.
There could be freshly exposed and newly formed minerals available for analysis and sampling around the impact crater, which would greatly enhance our understanding of the lunar impact process.
The products from a single impact may include shocked rocks, impact melt rocks, impact breccias, impact regolith, and shock lithified impact regolith \cite{osinski2023lunar}, allowing the collection of diverse sample types in a single mission.
Although the impactor would experience extreme shock during the impact process, some remnants may be left behind in the crater \cite{yue2013projectile}, which may provide additional information about the near-Earth impactor population.
As shown in Figure~\ref{fig:2}, a 10-m asteroid impactor would excavate a crater with a depth of $\sim$40~m, thus uncovering the upper megaregolith layer of the lunar crust.
For comparison, the deepest drill samples were taken only about 3 meters below the lunar surface \cite{allton1980depth}, highlighting the unique opportunity to access fresh, deep materials exposed by the impact events.
To reach the crater floor, the robotic rover must travel through the crater walls with a typical slope of up to 30$^\circ$ \cite{mahanti2014deep}, which may raise some engineering challenges in the future.
These in-situ investigations and samples would be helpful to understand 1) the lunar impact products and the impactor remnants, 2) the lunar subsurface layers and compositions, and 3) the shallow crust of the Moon and how it evolved after the lunar magma ocean \cite{elkins2011lunar}.

\subsection{Summary}

This paper provides an overview of the asteroid impact process on the Moon, which includes the asteroid impactor flux and the impact-induced craters, flashes, ejecta, and moonquakes.
The impact flux models predict one $>$10~m lunar impactor centennially, with the 60-m asteroid 2024~YR4 a possible candidate to hit the Moon in 2032.
For such impact energy scales, we use theoretical and empirical models to evaluate their impact results.
These scenarios motivate further scientific investigations of the dynamic response of the Moon in real time---an endeavor that would be unprecedented in lunar and planetary science.
We propose a combination of ground- and space-based telescopes, lunar-based seismometers and rovers, and in-situ investigations, which necessitate international collaborations.
Such efforts would significantly advance our understanding of the Moon's subsurface and internal structures, as well as its evolution history.

%%%%%%%%%%%%%%%% REFERENCES %%%%%%%%%%%%%%%

% The list of references goes after the main text and before the acknowledgements
% When preparing an initial submission, we recommend you use BibTeX, like this:
%
%\bibliography{science_template} % for a file named science_template.bib

%%%%%%%%%%%%%%%% ACKNOWLEDGEMENTS %%%%%%%%%%%%%%%

\section*{Acknowledgments}

YJ acknowledges the support of the National Natural Science Foundation of China (123B2038) and the Youth Talent Support Program of China Association for Science and Technology. BC is sponsored by Beijing Nova Program (20250484831). This work is also supported by the National Natural Science Foundation of China under Grant Nos. 62227901, U24B2048, and the national level fund No. KJSP2023020301.

\end{document}